\documentclass[fleqn,10pt]{wlscirep}
\usepackage[utf8]{inputenc}
\usepackage{mathtools}
\usepackage{csquotes}
\usepackage{mathdots}
\usepackage[T1]{fontenc}
\usepackage{bm}
\usepackage{float}
\usepackage[backend=biber,style=ieee]{biblatex}
\bibliography{bib} 
\usepackage{xcolor}
\usepackage[font={footnotesize}]{caption}

\title{Mechanisms of Resistive Switching in 2D Monolayer and Multilayer Materials}
\author[1]{M. Kaniselvan}
\author[2]{Y. R. Jeon}
\author[1]{M. Mladenović}
\author[1]{M. Luisier}
\author[2]{D. Akinwande}

\affil[1]{ETH Zurich, Department of Electrical Engineering and Information Technology, Zurich, Switzerland}
\affil[2]{The University of Texas at Austin, Department of Electrical and Computer Engineering, Austin TX, USA}

\begin{document} 
\flushbottom
\maketitle
\thispagestyle{empty}

\section*{Abstract}

The power and energy consumption of resistive switching devices can be lowered by reducing their active layer dimensions. Efforts to push this low-energy switching property to its limits have led to the investigation of active regions made with two-dimensional layered materials (2DLM). Despite their small dimensions, 2DLM exhibit a rich variety of switching mechanisms, each involving different types of atomic structure reconfigurations. In this review, we highlight and classify the mechanisms of resistive switching in mono and bulk 2DLM, with a subsequent focus on those occurring in a monolayer and/or localized to point defects in the crystalline sheet. We discuss the complex energetics involved in these fundamentally defect-assisted processes, including the co-existence of multiple mechanisms and influence of the contacts used. Examining the highly localized `atomristor'-type switching, we provide insights into the atomic motions and electronic transport across the metal-2D interfaces underlying their operation. Finally, we present the progress and our perspective on the challenges associated with the development of 2D resistive switching devices. Promising application areas and material systems are identified and suggested for further research.

\section{Introduction}

Resistive switching (RS) devices encode information through programmable changes in resistance. Their low operating energy and long retention times motivate their use in several applications, such as non-volatile storage, logic , radio-frequency communication, and as building blocks of new computing hardware \cite{Ielmini2018, Wu2018, Kim2024}. Unlike conventional semiconductor components, RS devices typically undergo modifications of their atomic structure during operation. Understanding the mechanisms of their operation - e.g., the connection between atomic structural changes, the subsequent alteration of electronic properties, and the resulting device-level behavior - is essential to develop viable devices and overcome the reliability limitations they currently suffer from. 

Typical RS devices are structured in a metal-insulator-metal (MIM) stack, where the insulating layers through which switching processes occur consist of oxide materials such as hafnium dioxide (HfO$_2$), tantalum pentoxide (Ta$_2$O$_5$), titanium dioxide (TiO$_2$), or solid electrolytes, for example germanium-antimony-tellurium (GeSbTe). So far, a rich variety of physical principles have been harnessed to realize reversible non-volatile resistance switching within active layers made with conventional materials, e.g., phase changes, control over ferroelectric polarization, or defect reconfiguration. However, ensuring that such changes occur reliably at low switching energies has proven challenging due to the stochastic nature of atomic movements and the relatively high energy required to alter atomic bonding environments. One way to achieve finer electrostatic control over the switching process consists of scaling down the device thickness - this allows for higher electric fields across the insulator, and reduces the distances required for the necessary atomic movements. However, this type of vertical scaling is limited by the existence of interface defect states and disorder in bulk materials. Exploring new crystalline materials is a viable route to partly overcome the limitations encountered when scaling down conventional oxides, and to push the energy-efficiency of RS device technology to its ultimate limits. 

One class of materials being scrutinized for nanoelectronic applications are 2D layered materials (2DLM). These materials are structurally-stable in atomically-thin layers, offering significant vertical scaling advantages. This property is already being leveraged for the development of logic devices, which can be scaled to a single atomic layer thickness without sacrificing carrier mobility \cite{Liu2021}. The past decade has witnessed significant advances in wafer-scale growth of 2DLM such as molybdenum disulfide (MoS$_2$), hexagonal boron nitride (hBN), and tungsten disulfide (WS$_2$). Motivated by their unique crystalline geometry, electrical properties, and increasing experimental maturity, researchers have uncovered multiple different RS phenomena within 2DLM, positioning them ideally for applications as non-volatile memory elements.

The development of 2DLM RS devices is currently in a nascent stage: The required stacks are often made with non-optimized or non-scalable deposition methods, most data is reported for a limited number of devices and switching cycles, and there is a lack of detailed information on yield and variability \cite{Lanza2018}. Nevertheless, remarkably high HRS/LRS ratios (I$_{\mathrm{LRS}}$/I$_{\mathrm{HRS}}$ as high as 10$^{11}$ \cite{Yang2023-2}), ultra-low switching energies (20 fJ \cite{Chen2020}), and new types of device functionality such as memtransistors \cite{Sangwan2018} have been demonstrated, making such atomically-thin RS devices, sometimes called atomristors \cite{Ge2017}, worth further exploration. 

In this Perspective, we examine the prospect of realizing ultra-scaled resistive switching devices with 2DLM active layers. The focus is set on the most common and technologically-relevant 2DLM whose properties have been investigated in multiple studies. We collect and present their mechanisms of RS, connecting experimental observations with theory and simulations where possible. We then use \textit{ab initio} simulations methods to highlight the complex energy landscapes involved during switching and resolve the combined influence of the contact and switching layer microstructures. Finally, we comment on the achievements so far at the material-, device-, and array-level, summarize the key advantages that 2DLM provide, and examine what remains to be addressed in the realization of atomically-thin RS devices. 

\section{Physics of resistive switching in 2DLM}

The operation of RS devices requires a mechanism by which the conductance of an active layer embedded between two electrodes can be reversibly tuned. The devices can then be switched between a High- (HRS) and Low-Resistance (LRS) State. Depending on the material stack, switching occurs either digitally (abrupt transition between the HRS and LRS) or in an analog way (gradual transition between several intermediate states). The resistance states can be non-volatile - e.g., retained in the absence of a bias - or volatile. Both cases result in a current vs. voltage (I-V) curve with a hysteresis. Transitions between the HRS and LRS can be bipolar, that is, voltages of opposite polarity are required for forward and reverse transitions, or uni/non-polar, where the same polarity is used for both transitions. For binary applications that only require two resistance states, the device performance is usually reported in terms of (1) the voltage required to transition to the LRS and HRS (the SET and RESET voltages) and (2) the ratio of conductance between the two states, the HRS/LRS ratio, also called dynamic range or memory window. These parameters, which define the device-level operating conditions, are visually summarized in \textbf{Fig. 1a-b}. 

Devices displaying this I-V characteristic are typically referred to as `memristors', or with different terms hinting at the physics behind their operating mechanisms. For example, Oxide-based Resistive Random Access Memories (OxRAM) are made of oxides in which the conductance can be tuned through the valence-change mechanism. In Conductive Bridging Random Access Memories (CBRAM) the resistance change depends on the growth/dissolution of filamentary paths made of metallic ions. In Phase Change Memories (PCM), switching results from reversible phase change of the material structure. In 2DLM switching devices, the active layer is a 2D multi- (\textbf{Fig. 1c}) or mono-layer (\textbf{Fig. 1d-e}). Many of the aforementioned physical effects have been reported for them as well. However, the specific mechanisms leading to RS in 2DLM are often difficult to resolve and may not generalize across devices made with the same material. They strongly depend on the growth and processing conditions of the 2D layer \cite{Huang2023, Lanza2018}, type of electrodes used \cite{Villena2019}, and biasing procedure \cite{Zhao2018-1, Li2018-8}. As the operating mechanisms in 2DLM-based memristors are still under investigation, here, we use the general term `resistive switching device' to encompass all such effects.

\subsection{Device architectures}

2DLM resistive switching devices can be realized either in vertical or lateral stacks. In the lateral case, the 2DLM is deposited directly on the substrate and well-separated electrodes are patterned on top or side, allowing for the current to flow laterally. This type of architecture has a lower integration density than the vertical one, but lends itself to the inclusion of a third, vertical control terminal to realize a field-effect. For example, due to the thinness of 2DLM channels, a gate-like terminal can exert enough electrostatic control to significantly modulate the current flow. Combining this gate-control with RS phenomena leads to so-called three-terminal `memtransistors', which have been implemented with MoS$_2$ \cite{Farronato2022, Sangwan2018}, WX$_2$ (X = S, Se) \cite{Leong2023}, and SnO$_2$ \cite{Huang2021}. Exposure of the active layer also enables optical control \cite{Ranganathan2020}. 

On the other hand, vertical configurations provide better lateral scaling, higher integration density, and finer control over the width of the active layer. In multilayer 2DLM devices, the thickness ranges from 2 to >10 atomic layers, but it can also be scaled down to the monolayer limit \cite{Bhattacharjee2020}. This level of scalability is enabled by the structural stability of individual monolayers, as well as by their crystalline structure. Conversely, bulk oxide materials are typically used in an amorphous phase which presents both dangling bonds (disordered surface states) and leakage through pinhole defects when scaled to thinner films. Remarkably, the area for switching can be localized to the site of atomic point-defects \cite{Hus2020, Nikam2021}. While single-point switching may be also achieved in bulk oxides by bridging a conductive filament, the variability of this process makes it difficult to precisely localize it using an applied bias. Resistive switching through atomic-scale conductive point(s) is thus unique to 2DLM.

\subsection{Identifying switching mechanisms}

Evidence for the switching mechanism is typically assembled from a combination of electrical and imaging measurements. Conductive atomic force microscopy (c-AFM) identifies nanoscale conductive paths by scanning the material surface. The presence of such conductive paths is indicative of filamentary mechanisms, in which the forming/breaking of a localized conductive pathway through the switching layer results in abrupt transitions between the HRS and LRS. Conductive filaments can result from metallic ions penetrating from one of the electrodes, or from percolating defect pathways. Electron energy loss spectroscopy (EELS) tracks the differences in spatial distribution of atomic species between different resistance states, providing further evidence about such atomic migration processes \cite{Pan2017}. Surface imaging reveals thermal plugs \cite{Datye2020} or physical extrusions at the interface caused by metal migration during conductive bridging \cite{Hus2020}, while bulk-imaging of device cross-sections using Focused Ion Beam (FIB) and Transmission Electron Microscopy (TEM) measurements directly capture conductive filamentary paths \cite{Chen2020} or phase changes \cite{Zhang2018}. In interface-type mechanisms, an applied bias modulates the density of interface defects at one of the electrode-active layer junctions. As the active layer is typically an insulating or semiconducting material, the density of defects effectively modulates the width of Schottky barriers to enhance or limit current flow. Interface-dependent phenomena are generally supported by a lack of clear conductive points, and dependence of the LRS resistance on device area \cite{Krishnaprasad2022}. In parallel with these characterization methods, material and device level modelling at \textit{ab initio} and atomistic levels of theory are used to help understand the operating mechanism(s) of a given device \cite{Huang2024}. 

This combination of imaging, electronic characterization, and simulation has so far revealed several distinct switching mechanisms in 2DLM. In lateral devices (\textbf{Fig. 1f}), RS has been seen to occur through modulation of grain boundaries bisecting the channel \cite{Sangwan2018}, cation migration from the electrodes \cite{Farronato2022}, or through electrostatic potential modulation resulting from the reversible migration of defects under the applied bias \cite{Li2018-8, Li2021-2, Leong2023}. In vertical 2DLM stacks (\textbf{Fig. 1g}), evidence for vacancy migration \cite{Yan2019, Tang2022}, Schottky-barrier modulation \cite{Krishnaprasad2022}, ferroelectric polarization \cite{Li2020, Yang2018, Hou2019}, phase change \cite{HeonLee2005, Zhang2018, Hou2023, Zhang2019}, and metal bridging \cite{Ge2020, Lu2021, Hus2020} have been observed. Focusing on 2D monolayers (\textbf{Fig. 1h}), switching has been suggested to occur via several different mechanisms, including single metal-atom (conductive point) adsorption \cite{Hus2020}, and localized Schottky-barrier modulation \cite{Lee2023}. 

As RS devices must block current flow in the HRS to achieve sufficiently high HRS/LRS ratios, the switching layers are required to have a finite bandgap. Unlike bulk semiconductors such as Silicon, many semiconducting 2DLM have sizable bandgaps (1.5 eV and more) and high effective masses ($>$ 0.5). 2DLM RS devices have thus been commonly demonstrated with both 2D insulators such as hBN, and 2D semiconducting transition metal dichalcogenides (TMDCs).

\subsubsection{Hexagonal boron nitride (hBN)}

Despite its simple atomic structure, hBN displays several interesting physical phenomena that can be exploited for RS. In many cases, the switching mechanism is attributed to the migration of contact atoms into the hBN layer, facilitated by the existence of grain boundaries and other intrinsic defects. This feature has been leveraged to realize crossbar arrays of hBN RS devices \cite{Xie2022, Wang2018, Chen2020}. \textbf{Figure 2a} shows a cross-section TEM image of a Ti/hBN/Cu device from Ref.~\cite{Chen2020}, indicating the locations of percolating conductive pathways through the multilayer. Several of these conductive points were found to simultaneously exist across the area of one device (\textbf{Fig. 2b}). Stacks of hBN also form energetically-favorable inter-layer defects, as predicted through density functional theory (DFT) calculations \cite{Strand2019} (\textbf{Fig. 2c}) and confirmed experimentally using TEM imaging \cite{Ranjan2023}. Electrical current calculations have shown that this effect is sufficient to modulate the conductance of the resulting RS device by a factor of 100, and is intrinsic to hBN stacks \cite{Ducry2022}. Such inter-layer defective pathways have been observed in devices with atomically-clean gold (Au)/hBN interfaces (\textbf{Fig. 2d-e}), where contact metal migration has a lower probability to take place \cite{Mao2022}. In other cases, they can facilitate the injection and transport of metallic ions to form filaments \cite{Jeong2020-2}.

Resistive switching is also found in monolayer hBN, which exhibits the ultimate thickness limit for vertical scaling (0.33 nm). HRS/LRS ratios of up to 10$^{11}$ have been measured in monolayer hBN RS devices \cite{Yang2023-2, Wu2019, Mao2022, Li2022-kk}. Switching mechanisms within the monolayer have not yet been clearly identified, but have been theoretically suggested to occur through the adsorption of contact atoms into individual Boron (or Nitrogen) vacancies \cite{Li2022, Wu2019, Li2022-kk}. Particularly remarkable is the fact that the HRS currents across the monolayer is low enough to sustain extremely high HRS/LRS ratios. Low tunneling currents in pristine (defect-free) hBN can be attributed to the high electron barrier height (3.07 eV \cite{Lee2011}) resulting from its large bandgap (6.0 eV \cite{Watanabe2004}). Current through hBN in the HRS is dominated by hopping-type tunneling transport through intrinsic defects \cite{Chandni2015, Wu2019, Mao2022}. The large HRS/LRS ratios observed may originate from the abrupt shift from defect-mediated hopping-transport to conduction through conductive points resulting from the diffusion of contact atoms into native defects.

\subsubsection{Transition metal dichalcogenides (TMDCs)}

Resistive switching has been reported in nearly a dozen different TMDCs. Due to its relative fabrication maturity as a semiconductor material, MoS$_2$ is one of the most well-explored members of the TMDC family, and the Au/MoS$_2$ interface has been particularly well investigated. In most of these devices, switching is attributed to Au migration, which can occur both within and between \cite{Ranganathan2020} MoS$_2$ layers. This mechanism has been observed on the smallest scale; in an MoS$_2$ monolayer contacted with an Au STM tip, the adsorption of a single Au atom into a sulfur vacancy led to a local decrease in resistance \cite{Hus2020}. This finding was substantiated with STM imaging (\textbf{Fig. 2f-g}) and measurements of a local physical protrusion at the adsorption site. The HRS/LRS ratios achieved by these single atom switching events are rather small (on the order of 10). However, macroscale studies using monolayer MoS$_2$ have reached ratios of several orders of magnitude \cite{Bhattacharjee2020}. These relatively large switching ratios likely originate from the presence of multiple switchable defect sites, the different geometry of extended planar electrodes compared to the STM tip, and/or the reduced contact resistance of macroscale devices. 

Metal migration belongs to the category of extrinsic switching mechanism and, as such, is most often observed in the presence of active electrodes. Similarly to the substitution of metal atoms into individual vacancies in monolayer structures, metal migration through multilayer TMDCs occurs preferentially through extended defects such as line-defects or grain boundaries. Some of the metal/multilayer TMDC stacks through which metal migration has been reported include Ti/PdSe$_2$ \cite{Li2021}, Ti/HfSe$_2$ \cite{Li2021-3}, Au/MoTe$_2$ \cite{Yang2023, Datye2020}, Ag/MoS$_2$ \cite{Farronato2022}, and Ag/SnS$_2$ \cite{Lu2021}. \textbf{Figure 2h-i} shows an example of Ag diffusion into a lateral Ag/MoS$_2$ device, imaged from above \cite{Farronato2022}. The presence of these varied combinations indicates that the TMDCs generally provide a fertile ground for the injection of metal ions.

Intrinsic RS mechanisms can also occur in TMDCs. For example, the migration of individual vacancies (\textbf{Fig. 2j}) can lead to significant resistance changes through a combination of filamentary, and non-filamentary (interface-type) effects. The filamentary mechanisms are driven by vacancy migration across multilayer TMDCs, typically guided by grain or flake boundaries \cite{Tang2022, Sangwan2015}, while non-filamentary switching can be induced by the accumulation of vacancies at one interface, thus modulating the metal-TMDC Schottky barrier height \cite{Li2018-8, Krishnaprasad2022}. RS can also be triggered by phase changes. Indium selenide (In$_2$Se$_3$) can undergo crystalline-crystalline phase changes with different resistivity \cite{Zhang2019, Ignacio2023}. Taking advantage of this effect, RS behavior has been demonstrated in In$_2$Se$_3$ thin films \cite{HeonLee2005}. Similarly, electric-field induced phase changes between the semi-metallic 1T' and semiconducting 2H phases have been observed in MoS$_2$ \cite{Lin2014, Zhu2018} and MoTe$_2$ \cite{He2022, Zhang2018, Hou2019, Hou2023}. Such phase transitions can be triggered by ion intercalation, e.g., Li \cite{Zhu2018}. Their occurrence is also made energetically favorable in the presence of point defects \cite{He2022}, external fields \cite{Zhang2022}, and mechanical strain \cite{Duerloo2014, Hou2019, Hou2023}. 

\section*{Box 1: Factors influencing the switching mechanism(s)}

It is crucial to fully understand the mechanisms underlying RS phenomena in order to engineer devices that meet specific functions and performance. As opposed to more conventional semiconductor devices, the operation of RS systems involves a close coupling between the movement of atoms and the flow of electrical current through the corresponding electronic structure of the material. In addition, existing observations show that device stacks made with the same 2DLM active layer can exhibit different driving mechanisms. These observations imply that the RS property and mechanisms are not only a function of the intrinsic material properties of the switching layer, but also depend on the 2DLM stoichiometry, its interfaces with electrodes, and the structure and nature of these electrodes.

\subsection*{Native defects}

In many cases, the existence of defects, either point- or extended- is central to the RS property. Pristine hBN was first explored as a promising dielectric. On the other end, defects in hBN allow for the emergence of RS phenomena with a large HRS/LRS ratio \cite{Shi2018}, albeit at the cost of increased leakage currents and reduced dielectric strength \cite{Shen2024, Knobloch2021}. A parallel can be made with the common dielectric material HfO$_2$, which when grown sub-stoichiometrically is one of the most technologically-mature RS materials. Such native defect distributions can be obtained by exploiting material growth methods which result in more defective films, such as atomic layer deposition (ALD) \cite{Wang2018}, chemical vapour deposition (CVD) \cite{Shi2018, Pan2017}, or metalorganic chemical vapour deposition (MOCVD) \cite{Kang2015}. Alternatively, defects can be added via post-processing methods such as annealing, electron-beam/ion/laser irradiation \cite{Li2021-2}, metal irradiation \cite{Pam2022}, or plasma treatment \cite{Li2018-8}. 

Filamentary RS devices typically require an electroforming step in which a high voltage is initially applied, after which the subsequent LRS and HRS transitions occur within a smaller operating voltage range. Many 2DLM devices display similar electroforming requirements, but several are also `forming-free', especially monolayer devices \cite{Ge2017, Ge2020}. For example, it was observed that Au-contacted multilayer hBN devices require an electroforming process when the hBN is exfoliated, but are forming-free with CVD-grown material \cite{Mao2022}. hBN devices, particularly multilayers, with relatively low defect densities require higher forming voltages to establish conductive filaments \cite{TejaNibhanupudi2024}. This suggests that the electroforming process effectively increases the defectiveness of the material to a level where the mechanisms behind SET and RESET transitions become energetically favorable. 

The existence of different defect distributions, such as point-defects or grain-boundaries, may also explain the coexistence of different switching phenomena within the same materials and across different studies. In exfoliated MoS$_2$ without any extended defects, conductive-point formation only occurs at single Sulfur vacancies \cite{Hus2020}. However in a more defective or polycrystalline material, these vacancies can be concentrated along grain boundaries \cite{Sangwan2015}. In MoTe$_2$, the existence of grain boundaries offers pathways for metal migration from the contacts \cite{Yang2023}, but point-defects (Te vacancies) lower the activation energy for the 2H-1T' phase transition \cite{He2022}.

\subsection*{Contacts}

While defects in the switching layer are often required to enable RS, the structural and electronic properties of the contacts play an equally important role. For example, the magnitude of defect-assisted transport depends on the relative alignment between the metal Fermi level and defect energy states \cite{Knobloch2022}; defect states which are outside of the Fermi window, defined by the electrodes of a metal-insulator-metal structure, do not significantly contribute to the current. Resistive switching is enabled if the SET transition alters the energetic distribution of defect states such that some become accessible within the Fermi window.

Devices in which intrinsic mechanisms are observed, such as defect-migration or phase transitions, typically include a combination of relatively inert contacts \cite{Li2021-3} and atomically-flat interfaces \cite{Mao2022}. Although these intrinsic mechanisms can occur in all cases, in the presence of `active' (low cohesive energy) contacts, the migration of extrinsic contact atoms is instead observed \cite{Shen2024}. This happens because resistive switching follows a positive feedback loop between current flow, Joule heating, and local electric field changes. The mechanism which is initially energetically favorable, and subsequently drives the most significant changes in conductance tends to dominate. Since the resistivity of conductive metal filaments is lower than that of the states created by vacancy defects, they can easily become the dominant factor. The switching energy and volatility of RS devices are also highly dependent on the contact metal used; in hBN, for example, non-volatile switching has been reported with Au \cite{Chen2020, Mao2022} and Ti \cite{Pan2017} electrodes, while Ag \cite{Chen2020}, Pt \cite{Chen2020}, and Ni \cite{Vlkel2023} also produce volatile or so-called `threshold switching'. A transition from threshold to non-volatile switching in the same device can be achieved by increasing the compliance current \cite{Shi2018} or by alloying different metals \cite{Passerini2023}.

Graphene contacts to 2DLM active layers are particularly interesting as they lead to atomically clean 2D-2D interfaces \cite{Ge2017, Zhang2018}. Thin layers of graphene have been used as a selective physical barrier to limit the diffusion of metal contact atoms into the active layers \cite{Zhu2019}. Beyond their structural advantages, graphene-insulator stacks also host a variety of interesting electronic phenomena. For example, resonant tunneling effects might emerge at biases which maximize the number of transport channels satisfying momentum-conservation. This effect can be engineered by varying the twist angle between the two graphene layers \cite{Mishchenko2014}, providing a tuning knob to suppress current and achieve a lower HRS.

\subsection*{Polarity}

RS devices can be either polarity-dependant (`bipolar') or independant (`unipolar'). The polarity-dependence of bipolar devices results from field-driven processes that move charged ions within the switching layer and drive atomic structural changes. The switching polarity is determined by the direction of the electric field applied during the electroforming process. Conductive filaments tend indeed to grow from one electrode interface to another and are asymmetric once formed, with a constriction near one end. An example of such an asymmetric conductive filament is pictured in \textbf{Fig. 1g} (metal filament). As a result, the filament dissolves more easily at one interface. In unipolar devices, however, the same voltage polarity is used to trigger both SET and RESET processes. In this case, the switching is likely driven by Joule heating generated within the 2DLM \cite{Yang2012}.

Devices based on filamentary growth across multiple layers are often bipolar. However, similarly shaped hBN stacks have been operated in bipolar \cite{Chen2020} and unipolar \cite{Mao2022} modes, while Ag/MoS$_2$/Au has shown both bipolar and unipolar switching within the same device, depending on initial voltage polarity \cite{Zhao2018-1}. This suggests that the switching polarity strongly depends on electrode environment and coexistence of field-driven and heat-induced mechanisms. In fact, many of the 2DLM RS monolayers demonstrated so far can be operated in a unipolar mode \cite{Ge2017, Li2022-kk, Wu2019}. In a study including several symmetrically Au-contacted TMDC monolayers, bipolar and unipolar switching have been found to occur in each configuration \cite{Ge2017}. The two modes of operation require a similar current level (above the compliance current used for the SET process) and magnitude of applied voltage to trigger the RESET process, implying that the bipolar and unipolar RESET in monolayer RS devices may rely on a common, thermally-activated and non-polar phenomenon.

\section{Mechanisms of switching within the monolayer}

Monolayer devices fully exploit the dimensional advantages provided by 2DLM, and serve as a test-bed for a deeper exploration of the mechanisms involved. Many resistive switching processes can be broken down into primitive events, each involving a form of atomic movement. The energies required to activate such events can be estimated with methods such as \textit{ab initio} Nudged Elastic Band (NEB) \cite{JNSSON1998}, which relies on DFT to compute energy landscapes and atomic motions through them. As a representative system for device-level RS investigations, we consider the interface of an Au electrode and monolayer MoS$_2$ active layer, where evidence of truly atomic point-switching has been demonstrated \cite{Hus2020}. The Au surface is typically found in the highly-packed (111) plane (\textbf{Fig. 3a}). In \textbf{Fig. 3b-d} we consider mechanisms that occur between this Au surface and the semiconducting 2H phase of MoS$_2$. The first two processes begin with the extraction of an Au atom from an atomically-flat contact, and end with the formation of an Au interstitial in the top Sulfur plane (\textbf{Fig. 3b}) or adsorption into an existing Sulfur vacancy (\textbf{Fig. 3c}). In the third process, we assume an Au adatom migrating across the interface and then adsorbing into a Sulfur vacancy nearby (\textbf{Fig. 3d}). The associated computational details can be found in the Supplementary Information. 

The formation of Au interstitial in the Sulfur plane is a highly unlikely process, requiring roughly 6 eV to be activated. Furthermore, the reverse process is barrier-free and can occur in the absence of added energy. This could explain why metal migration-induced switching is typically not seen in pristine materials. The presence of a Sulfur vacancy adsorption site, however, lowers the activation energy to 0.21 eV. The process appears non-volatile, with nearly identical energetic barriers for the forward and reverse processes. When the electrode surface is not atomically smooth but `rough' and contains diffusing adatoms, switching may occur from the adsorption and release of adatoms rather than electrode atoms. In the case of the Au(111) surface, the activation barrier for Au adatom diffusion across the surface is as low as 0.12 eV \cite{Bon2019}, rendering them mobile. In the case of such `rough' electrodes, adatoms have fewer bonds with other Au atoms, and are less stabilized by the Au surface along their trajectory into an adsorption site. This leads to a slightly higher activation energy of 0.34 eV for this process. In general, the mechanisms involving defective MoS$_2$ are activated at similarly low energies and may coexist across the interface. 

In \textbf{Fig. 3e-f} we extend the investigation of the lowest-energy adsorption event from \textbf{Fig. 3b} to the Au(100) surface plane, and Ag, which is another common electrode for 2D RS devices. The Au(100) plane requires a slightly higher energy to dissociate atoms from the surface. This may be due to the reduced surface packing density, which results in lower stabilization of the diffusing atom and necessitates a larger rearrangement of the Au surface around the newly formed vacancy. Unlike adsorbed Au atoms, which are stable in their final position and face an energy barrier to return to the electrode, Ag adsorbents have a nearly barrier-free reverse transition, which could explain why volatile switching is also seen with this electrode material when interfaced with MoS$_2$.

The atomic structures of the HRS and LRS enable different mechanisms of electronic transport, resulting in distinct conductance states. We illustrate this in \textbf{Fig. 3g-i} through energy band-diagrams. Across a pristine Au-monolayer MoS$_2$ interface, transport can occur through direct tunneling or Schottky emission above the barrier. The existence of native defects in the HRS makes these situations less likely to be found in RS devices. Rather, defects such as vacancies create in-gap states supporting additional trap-assisted mechanisms. A combination of different transport events, such as tunnelling into a defect site followed by a subsequent phonon-assisted emission process, is also possible. This picture is consistent with the non-linear I-V characteristics typically observed in the HRS and increased current at higher temperatures \cite{Ge2017, Wu2019}, which allow for carriers to be injected at higher energies. We note that the barrier heights are influenced by the locations of defect states in the bandgap due to the possibility of strong Fermi level pinning at metal-2D interfaces \cite{Kim2017}. Conversely, the LRS typically displays highly linear I-V characteristics and is degraded at increased temperatures \cite{Ge2017, Wu2019}, a typical signature of Ohmic transport (\textbf{Fig. 3i}). This physical picture also implies that methods to engineer higher tunneling effective masses, such as strain-engineering, or defects located deeper in the bandgap can suppress the HRS conductance and achieve higher HRS/LRS ratios.

\section{State-of-the-art and current progress}

Differences in material quality, device fabrication, and measurement procedures among research groups complicate the extraction of high-level trends from published data. Besides the 2DLM layer thickness, defect density, contact metal, and interface quality, there are also significant variations in the device dimensions and post-fabrication treatments such as annealing. Regarding the measurement procedure, the HRS/LRS ratio is a strong function of the chosen compliance current (I$_{CC}$), which is imposed to prevent device breakdown. 

Despite these differences, it is worth tabulating the demonstrated performance so far to provide an overview of the state-of-the-art. In \textbf{Fig. 4a-c} we summarize key device metrics - SET voltage, HRS/LRS current ratio, and HRS/leakage current - for several monolayer (\textbf{Fig. 4a-b}) and multilayer (\textbf{Fig. 4c}) 2DLM RS devices reported so far. These metrics serve as proxies for the power dissipation and energy-efficiency prospects of 2DLM RS devices. The currents determine the leakage levels during readout operations, while the voltages indicate the fields required to activate the relevant RS mechanisms. We note a trade off between the SET voltage measured and the sweep rate used, so that the correlation to switching energy is approximate. A few works have directly calculated the switching energy based on the time (t$_{\mathrm{SET}}$) required for the LRS transition to occur, reporting ultra-low switching energies of 100 fJ (SnS \cite{Lu2021}, metal-filament), 20 fJ (hBN \cite{Chen2020}, metal filament) and 150 aJ (MoSe$_2$ \cite{Hou2023}, phase-change).

For monolayer devices, the HRS/LRS ratio is often limited by the HRS leakage current (\textbf{Fig. 4b}). As a result, higher bandgaps can result in lower HRS currents and higher HRS/LRS ratios. So far, monolayer hBN with Ag electrodes achieves the highest HRS/LRS ratio of \(\sim \)10$^{11}$ with a low switching voltage below 0.5 V \cite{Yang2023-2}. In the case of multilayer devices (\textbf{Fig. 4c}), the SET voltage and achievable HRS/LRS ratio are highly influenced by the device thickness. Higher ratios are also obtained with thicker multilayers which more efficiently block leakage currents. It is worth noting that the distribution of HRS/LRS ratios achieved by multilayer devices of various thicknesses is still within the same range as that of their monolayer counterparts. This may indicate that the `effective' switchable layer after the initial electroforming process is limited to a single/few layer(s), even in multilayer active regions \cite{Huang2023}. The extra layers, however, can laterally confine the conductive filament growth and suppress direct tunneling leakage currents.

Array-level demonstrations have explored whether the theoretical high-integration density of 2DLMs can be achieved in practice. Crossbar arrays of 2DLM have been reported for multilayer hBN (10$\times$10 devices \cite{Chen2020}), multilayer MoTe$_2$ (5$\times$5 devices \cite{Yang2023}), and multilayer SnS (32$\times$32 devices \cite{Lu2021}) with yields of 98\%, 83.7\%, and nearly 100\%, respectively. The integration density can be further improved through 3D stacking, as demonstrated for TMDC RS devices \cite{Tang2022}.

At the current state, it remains unclear how the performance of 2DLM devices compare quantitatively against those of conventional oxides. Such a comparison is difficult to make due to the differences in maturity between these two technologies, and the limited data and statistics available for the nascent 2DLM devices compared to the more mature oxide-based RRAM.

\subsection{Trends in scaling and variability}

The stochastic nature of individual ionic movements translates into a certain level of cycle-to-cycle and device-to-device variability. Device-to-device variability in multilayer 2DLM RS devices has been found comparable to that of binary oxides, as measured in Ref.~\cite{Chen2020} across devices belonging to the same multilayer hBN crossbar array (\textbf{Fig. 4d}). The physical origins of this variability are attributed to fluctuating layer thicknesses, for example between grains \cite{Villena2019} or substrate terraces \cite{Shi2021-2}. Various methods have been shown to reduce it, including the use of atomically-clean substrates \cite{Mao2022}, or guiding filament formation through grain boundaries \cite{Li2021, Yang2023}. Even in devices with structurally very homogenous active layers, slight differences in thermal, electrical, and energetic environments may lead to significant cycle-to-cycle variability. For example, a difference in SET voltage was observed even when switching single defects at two different locations of the same exfoliated MoS$_2$ flake \cite{Hus2020}. Additional cycle-to-cycle variability in monolayer devices with extended contacts may emerge if a different set of conductive points is created each time, each with a slightly different switching environment. 

To investigate whether the switching energy can be reduced and the HRS/LRS ratio maintained with decreasing device dimensions, we show in \textbf{Fig. 4e-f} published trends in thickness/area scaling. All data points within each of these subplots are from the same report, such that the extraction of trends is possible. The SET voltage is reduced for thinner active layers (\textbf{Fig. 4e}) due to higher electric fields. HRS/LRS ratios are generally, but not always, degraded by vertical scaling (especially in few layers), and typically improve with lateral scaling \cite{Yang2023, Ge2017} (\textbf{Fig. 4f}). The former effect results from increased leakage currents across thinner 2DLM. The latter emerges from the localized nature of these conductive points; the LRS resistance is usually nearly independent of the device area, while the HRS resistance increases for smaller device area \cite{Shi2017, Ge2017, Xie2022}, leading to an effective increase in dynamic range. In filamentary devices, the trend in SET voltages when scaling down device area is counter-intuitive; higher voltages are required for lower device areas \cite{Zhu2019, Shi2017}. This effect can be seen in the inset of \textbf{Fig. 4f} for a multilayer hbN device, and has also been observed in RS devices made with binary oxide active layers and relying on a similar filamentary mechanism \cite{Chen2013}. It results from the preferential nucleation of conductive points at the electrically weakest locations. Larger device areas are likely to sample more inhomogenous regions of the active layer, leading to lower-energy nucleation sites for conductive filaments. 

\section{Outlook: Resistive switching in 2DLM - why, how, and where?}

Driven by logic applications, the 2DLM technology is experiencing rapid progresses and its experimental maturity keeps increasing. $n$-type transistors made with 2DLM channels are currently undergoing lab-to-fab developments. Contemporary research is tackling remaining limiting factors such as the realization of $p$-type transistors with high mobility, the reduction of contact resistances, the development of efficient doping techniques, or the creation of clean interfaces with dielectrics. Low-temperature or large-area growth combined with suitable transfer approaches offer a path towards the integration of 2DLM on existing CMOS platforms, which remains one of the key challenges for future development. We can thus anticipate a possible transition towards new all-2D systems, as a means to overcome the limitations posed by bulk materials at the nanoscale. This also concerns RS devices, whose research has so far been dominated by metal oxides (HfO$_2$, TiO$_2$, Ta$_2$O$_5$, WO$_3$,...). The situation is changing with the emergence of 2DLM and their multiple resistive switching mechanisms. Nevertheless, to be viable for memory or in-memory computing, 2DLM should offer a competitive advantage over more conventional oxides in terms of energy/power performance, programmability, scalability, and/or functionality. In other words, their use as RS devices needs a compelling argument.

\subsection{Why: Structural advantages, diverse materials and mechanisms}

The incorporation of 2DLM as an active layer to build RS devices has several advantages. First, the ability to down-scale the device dimensions, particularly the switching layer thickness, unlocks the possibility of low power and energy operation. The latter are crucial features of non-volatile RS devices. While metal-oxide RS devices of 4 nm active layer thickness can show a HRS-to-LRS switching power of about 230 nW  (I$_{\mathrm{CC}}$ = 50 nA, V$_{\mathrm{SET}}$ = 4.6 V) \cite{Pi2018}, the same metric in hBN devices with an active layer thickness of roughly 2 nm does not exceed the fW range (I$_{CC}$ = 110 fA, V$_{SET}$ = 0.4 V) \cite{Chen2020}. Secondly, a large palette of methods is available to control their atomic and electronic structures, for example through strain, twist-engineering, defect-engineering, or Fermi level depinning \cite{Kim2017}. All these techniques can be exploited as tuning knobs to improve the RS mechanism. 

Furthermore, many RS mechanisms are fundamentally defect-mediated, and achieving reproducible device operation becomes a question of defect controllability. The energy to create chalcogen vacancies in MoS$_2$, for example, is fairly low \cite{Komsa2012}. As the majority of performance degradation factors in RS devices can be attributed to uncontrollable or undesired atomic re-locations, a material system in which structural changes can be optimally manipulated is advantageous. The simpler atomic structures of 2DLM devices allow for more precise defect engineering approaches to be employed. Moreover, their exposed surfaces are accessible to a variety of characterization tools to investigate their operation and variability. Finally, active layers based on 2DLM have the potential to minimize the influence of some practical considerations such as variability. When scaled down to the mono- or few-layer, 2DLM consist almost entirely of their surfaces, thus exhibiting a smaller configuration space of possible atomic rearrangements. In the ultimate limit of point-switching, this surface variability would also be reduced due to localization of atomic dynamics to a few point sites.

The prospect of single-atom switching in 2DLM monolayers represents the ultimate limit of miniaturization and a significant opportunity towards reaching energy efficiency near the Landauer limit, the theoretical thermodynamic minimum energy required to toggle a bit, \(\sim \)3 zeptojoules, \cite{Landauer1961}. Single-atom switching involves the movement and adsorption of an individual atom or ion from the electrodes into a defect in the 2DLM in order to modulate the resistance state \cite{Hus2020}. As such, this mechanism is expected to consume a very low energy per switching. Moreover, it should also afford extremely fast switching speeds since the length scale of atom/ion movement for a switching event in monolayers is of the order of the van der Waals gap, 1 nm. In practise, the motion of few atoms might be needed to ensure high HRS-to-LRS ratios. Still, research is needed to realize reliable and reproducible single- to few-atom switching at the device level.

Considering the current state of research, several specific materials and mechanisms in 2DLM have emerged as especially promising. MoS$_2$ and hBN are the most mature 2DLM, and present the lowest barrier-to-entry for mass production. The proven ability to down-scale these materials to the monolayer allows to fully exploit their layered nature and to achieve sufficiently high HRS/LRS ratios because of the electronic current-blocking properties of a crystalline monolayer. The filamentary switching mechanisms observed in them can be highly localized, offering the possibility of increased integration density. The laterally-confined nature of the current flow may also lead to less thermal and electrical cross-talk between devices in a crossbar array. 2DLM phase-change memories are another promising technology. The phase-change mechanism is well-established in bulk RS components \cite{Optane}. However, the retention of these devices, typically made of GeSbTe compounds, is limited by resistance drift in the amorphous state corresponding to the HRS. This drift has been attributed to different structural configurations of the amorphous phase, and their relaxation over time and temperature \cite{LeGallo2020}. In 2DLM such as In$_2$Se$_3$ \cite{Zhang2019} and MoTe$_2$ \cite{Hou2019}, the phase transition occurs between crystalline phases rather than across a crystalline/amorphous boundary, a property which could result in more predictable resistance values with less drift and improved retention times. Future work in this area is nevertheless required to determine whether sufficient resistance ratios can be retained when the 2DLM thickness is scaled to the monolayer.

\subsection{How: Guidelines for further development}

Advancing the field requires going beyond demonstrations of RS in prototype devices: the conditions under which switching occurs (specific growth, process, contacting, biasing...) should be thoroughly investigated such that they can be reproduced reliably. Key performance metrics, such as endurance, variability, and reliability, should be reported along with I-V curves, in line with best practices advocated in Ref.~\cite{Lanza2018}. On the materials level, a deeper understanding of the switching mechanisms is crucial. This can be partly achieved through computational approaches where the devices under test are described at atomistic resolution and their combinatorial design space are explored `in silico'. \textit{Ab initio} modelling appears as an ideal tool to accompany the experimental development of 2DLM RS devices and predict their materials and operational properties. Due to the reduced dimensions of 2DLM, realistic device structures can be constructed atom by atom, similar to the ones fabricated in laboratories. Such a step is rather challenging for traditional oxide-based systems. Hence, 2DLM offer unprecedented opportunities for rational device design based on material predictions from computational electrochemistry. 

In addition, the kinetics of switching should be controlled more precisely. So far, defects have typically been introduced indiscriminately in order to observe or enhance RS. Increasing the defectiveness of the material, however, results in trade-offs between the achievable memory window and device-to-device variability. An initially defective material is required to sustain the possibility of switching events and to boost the LRS current, but it can simultaneously increase the leakage current in the HRS. Defects generated during material growth are randomly distributed, and conductive paths form preferentially in highly defective areas. Hence, on an array level, a device occupying a more defect-rich area of the active layer may have a higher HRS current than its neighbors, leading to increased device-to-device variability. The latter is also exacerbated as the devices are scaled down, as devices with a small active area are more likely to sample regions with different defect concentrations. Monolayer 2DLM grown in large areas can present numerous extended defects, which make single conductive-point switching more difficult to achieve than in exfoliated crystals. In addition, when the defect density is quite high, individual defects can cluster to form larger voids potentially leading to an irreversible shorting of the vertical structure \cite{Wu2022}. 

Solutions have been proposed in the form of defect-engineering. For example, selective ion-irradiation can be applied to the pristine 2DLM to carve out preferential RS sites. Ref.~\cite{Jadwiszczak2019} demonstrated selective defect-engineering in lateral MoS$_2$ memtransistors, where the artificial placement of a defect line bisecting the active area localizes the RS events to achieve higher reliability. Designing such an approach for vertical devices could increase the proportion of defects that contribute to switching, reducing those which only induce leakage in the HRS. Downscaling also has advantages in this regard; by reducing the cell area below the grain size, it is feasible to realize RS devices that are on average confined to a single crystal domain with point-defects in an otherwise polycrystalline 2D film. This effect is leveraged in contemporary studies reporting resistance switching at conductive-points \cite{Kim2022-2}. The physical properties of specific defect types can be explored using high-throughput computational approaches \cite{Davidsson2023}. Further research into defect identification and controlled defect placements could thus improve the reproducibility of switching events.

\subsection{Where: Application areas}

We see 2DLM well-positioned for several emerging memory applications. The market for conventional non-volatile memory applications is currently dominated by V-NAND (vertical NAND) flash where the reduced planar integration density is compensated by stacking cells in the vertical dimension. However, the stacking ability of this technology is reaching its limits, and its power consumption is high due to the large voltages required \cite{Yoon2019}. As a result, there is now an incentive to develop resistive alternatives, with a higher integration density enabled by specific devices. From this viewpoint, there is an attractive prospect of developing ultra-high density, defect-engineered 2DLM RS arrays with single defects localized at every crosspoint. Such an arrangement would result in uniform switching properties, unlocking the promising traits of ReRAM (area scaling, energetics, retention, reliability, etc) at reduced variability.

The recent interest in developing synaptic devices, which present not only an HRS and LRS, but ideally a continuum of states, is also being explored with 2DLM active layers. In such small 2DLM active areas, analog and digital switching likely share common mechanisms. The presence of intermediate states may be caused by the finer modulation of filaments within multilayers, or by the activation of subsets of discrete conductive points across monolayers. This continuum of states and the more gradual variations between them makes such configurations promising as Multi-Level Cells (MLCs) or as building blocks of emerging neuromorphic computing architectures \cite{Yang2012}. Also, the larger surface/bulk ratio possible in lateral device configurations enables new compute functionality with a third, vertical terminal. Finally, several other applications for 2DLM RS devices have emerged in recent years, including their use as radio-frequency switches \cite{Kim2024}, which takes advantage of their very low LRS, $<$ 10 $\Omega$. With further development, we envision that 2DLM devices will provide a versatile platform for storage, processing, and communication of information on a truly atomic level.

\section{Acknowledgements}

M.K., M.M., and M.L. acknowledge funding from the ALMOND SNSF Sinergia project, (grant no. 198612), the NCCR MARVEL (grant no. 205602), and the Werner Siemens Stiftung Center for Single Atom Electronics and Photonics. They also thank the Swiss National Supercomputing Center (CSCS) under project s1119 for computational resources. M.K. acknowledges the Natural Sciences and Engineering Research Council of Canada (NSERC) Postgraduate Scholarship (PGS-D3). D.A. acknowledges the support of the National Science Foundation (NSF) grant no. 2422934 and the Office of Naval Research (ONR) grant N00014-24-1-2080. In addition, Y.R.J. acknowledges the support of Samsung.

This is the accepted manuscript of an article published by Nature Materials. The published version is available online at: https://www.nature.com/articles/s41563-025-02170-5.

\section{Author Contributions}

D.A and M.L initiated and supervised the preparation of this manuscript. M.K led the writing and design of figures, with inputs and discussion from all authors. Y.R.J collected data for Fig. 4, and M.M performed the NEB simulations in Fig. 3. 

\section{Competing Interests}

The authors declare no competing interests.

\newpage
\printbibliography
\newpage
\begin{figure*}[h]

\centering\includegraphics[width=\textwidth]{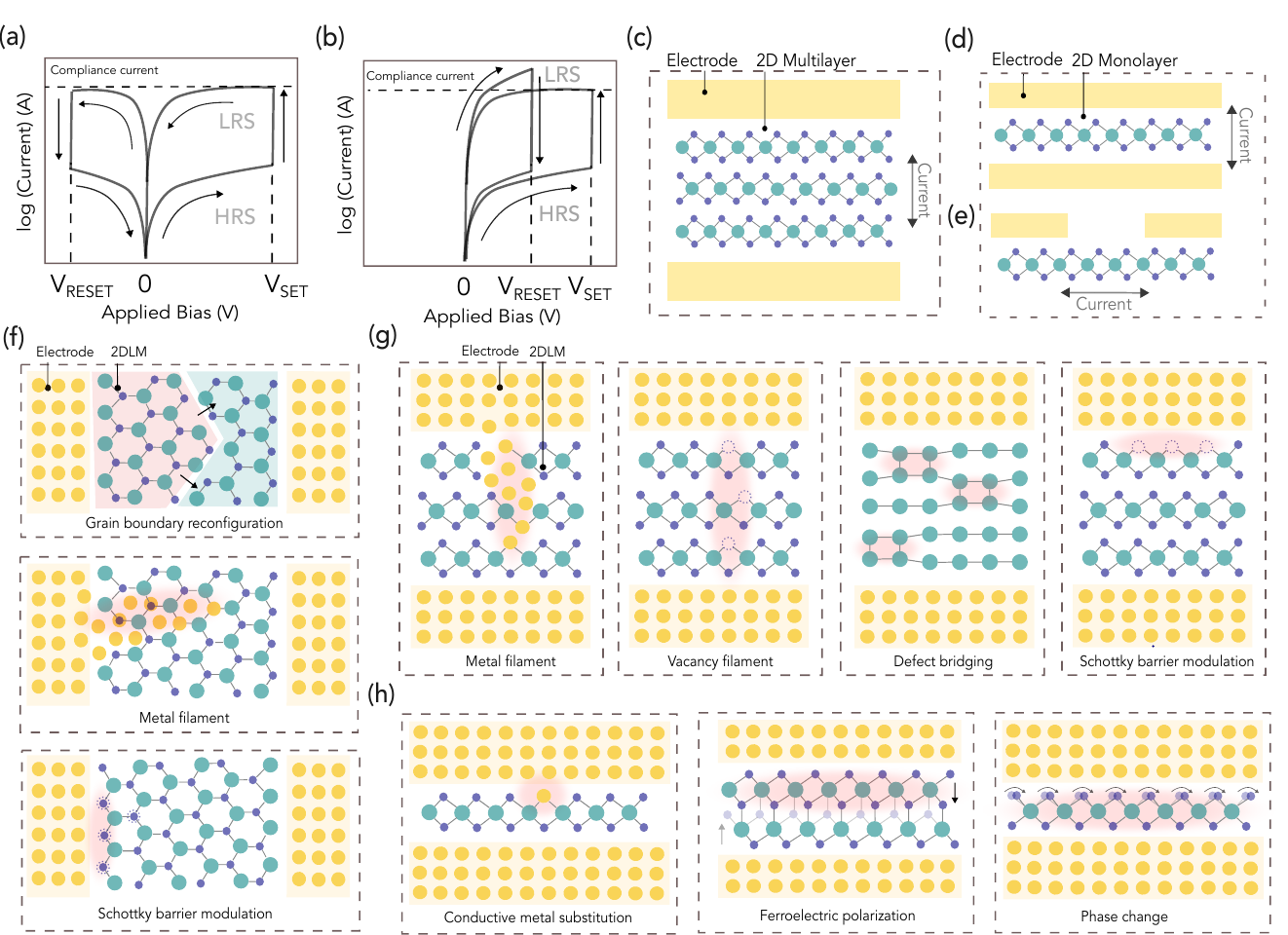}\par
  \caption{\textbf{Operation, structure, and mechanisms of 2DLM RS devices}. A schematic pinched-hysteresis current vs. voltage (I-V) curve resulting from resistive switching (RS) device measurement is shown in \textbf{(a)}/\textbf{(b)} for bi/nonpolar switching characteristics, indicating the non-volatile transitions from the High Resistance State (HRS) to the Low Resistance State (LRS) and back. The device structure usually consists of a switching layer sandwiched between two contacts. The switching layer can be realized with 2DLM in \textbf{(c)} multilayer or \textbf{(d-e)} monolayer forms, and in \textbf{(d)} vertical or \textbf{(e)} lateral configurations. The black arrows in each case indicate the direction of current flow across the 2DLM. Schematics of the switching mechanisms are shown in \textbf{(f)} lateral and \textbf{(g-h)} vertical 2DLM devices. The mechanisms are sorted by their scaling potential, both vertical (limited to multilayer, or capable of existing in monolayers) and lateral (requires extended area for switching, or can be localized to a point). The light shaded areas highlight the conductive areas once the device is switched to its LRS. In all cases, the 2DLM has the structure of a transition metal dichalcogenide or hexagonal boron nitride, and is representative of diverse 2DLM. The mechanisms shown have been reported experimentally, for example in Refs (f - grain boundary reconfiguration) \cite{Sangwan2018}, (f - metal filament) \cite{Farronato2022}, (f - schottky barrier modulation) \cite{Huang2021}, (g - metal filament) \cite{Chen2020}, (g - vacancy filament) \cite{Yan2019}, (g - defect bridging) \cite{Ducry2022, Mao2022}, (g - Schottky-barrier modulation) \cite{Pam2022}, (h - conductive metal substitution) \cite{Hus2020}, (h -  ferroelectric polarization) \cite{Li2020}, (h - phase change) \cite{Hou2019} ).}
  
\end{figure*}

\newpage
\begin{figure*}[h]

\centering\includegraphics[width=\textwidth]{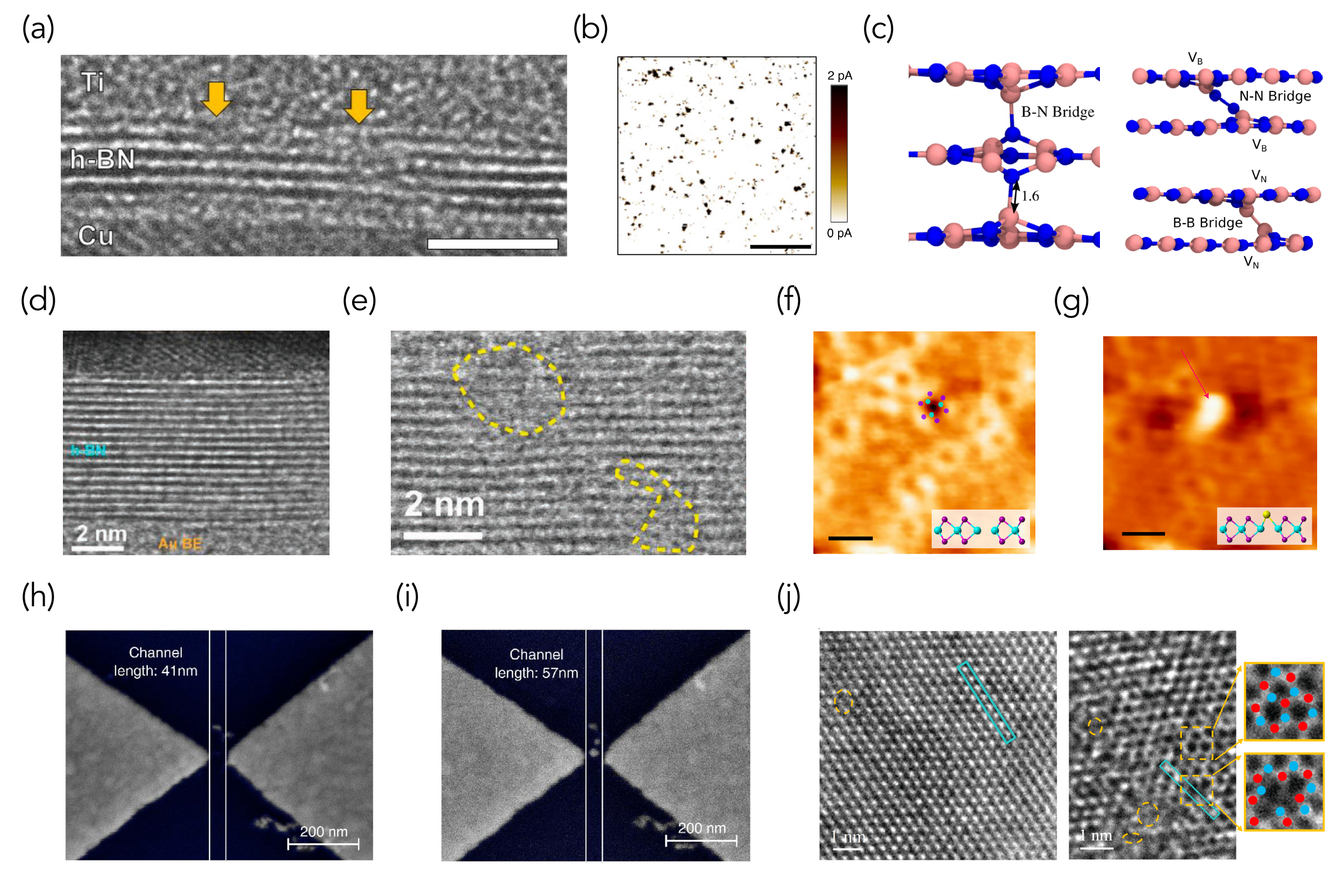}\par
  \caption{\textbf{Structural transitions leading to resistive switching in hBN and TMDC devices}. (\textbf{a}) Cross-sectional TEM image of a Ti/hBN/Cu device, indicating the location of structural distortions in the hBN multilayer. The scale bar is 1.5 nm. A conductive AFM map of the surface of this structure is shown in (\textbf{b}) (scale bar is 250 nm). (a)-(b) are adapted from Ref \cite{Chen2020}. (\textbf{c}) Molecular structures of different interlayer bridges formed across hBN layers, taken from Ref \cite{Strand2019}. Similar bridge structures have been observed in (\textbf{d}) TEM images of hBN RS devices with atomically flat contacts, as highlighted by the yellow dashed lines in (\textbf{e}) after switching to the LRS (adapted from Ref \cite{Mao2022}). (\textbf{f})-(\textbf{g}) Scanning tunnelling microscope (STM) images of a RS mechanism in which a single Au atom from the STM tip substitutes into a sulfur divacancy in an MoS$_2$ monolayer, leading to (\textbf{f}) HRS and (\textbf{g}) LRS states, adapted from Ref \cite{Hus2020}. (\textbf{h})-(\textbf{i}) shows the existence of Ag contact migration into the active layer of a lateral MoS$_2$ device (\textbf{h}) before and (\textbf{i}) after switching, adapted from Ref \cite{Farronato2022}. (\textbf{j}) TEM imaging of tungsten and sulfur vacancies in WS$_2$, adapted from Ref \cite{Yan2019}.
  }
\end{figure*}

\newpage
\begin{figure*}[h]

\centering\includegraphics[width=\textwidth]{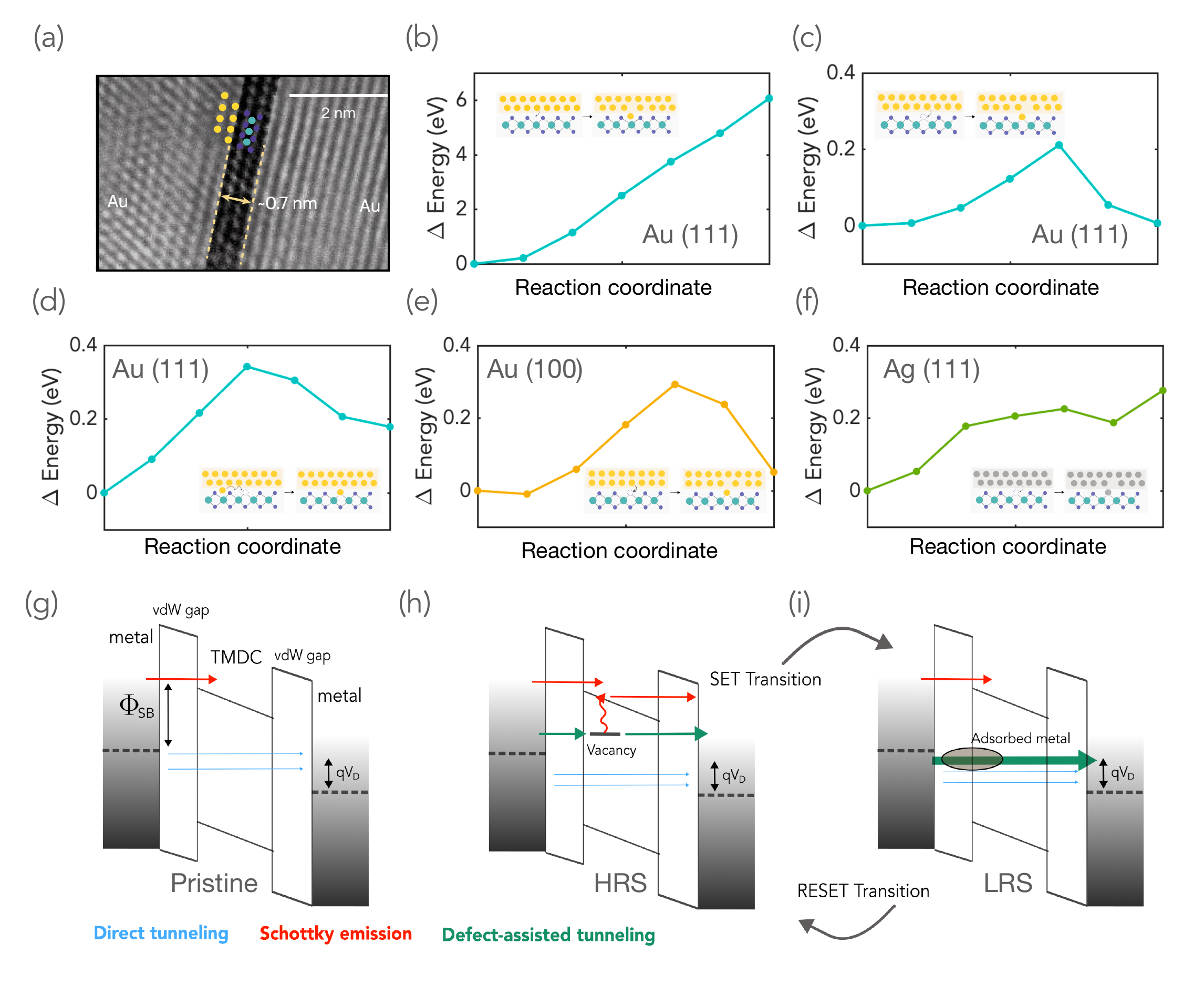}\par
  \caption{\textbf{Resistive switching mechanisms at the electrode/monolayer 2H-MoS$_2$ interface}. \textbf{(a)} TEM image of the Au-2H MoS$_2$ interface in which the Au (111) surface termination is clearly visible, adapted from Ref \cite{Ge2017}. Schematics representing \textbf{(b)} Au metal interstitial creation, \textbf{(c)} Au adsorption into a Sulfur vacancy site, and \textbf{(d)} adsorption of an Au adatom at the interface into a Sulfur vacancy site are shown on top of plots representing the energy surfaces along the minimal-energy pathways for each of these processes. An \textit{ab initio} Nudged Elastic Band (NEB) approach was used for that purpose \cite{JNSSON1998}, considering the (111) crystallographic plane of Au. In \textbf{(e-f)} the mechanism of metal adsorption from \textbf{(c)} is further explored for \textbf{(e)} the (100) crystallographic plane of Au and \textbf{(f)} the (111) plane of Ag. In each case, the reaction coordinate enumerates seven states found during the simulation of the minimum-energy trajectory. All calculations are performed with initial and final images corresponding to the shortest path available for the mobile species to complete such a transition. \textbf{(g-i)} Electronic transport mechanisms across the vertical Au/monolayer-MoS$_2$/Au stack in case of \textbf{(g)} a pristine monolayer, \textbf{(h)} a monolayer with intrinsic defects, such as Sulfur vacancies, and \textbf{(i)} a monolayer with an adsorbed Au atom. The wavy arrow indicates a phonon-assisted transition to a higher energy level, leading to enhanced Schottky emission.}
\end{figure*}

\newpage
\begin{figure*}[h]

\centering\includegraphics[width=\textwidth]{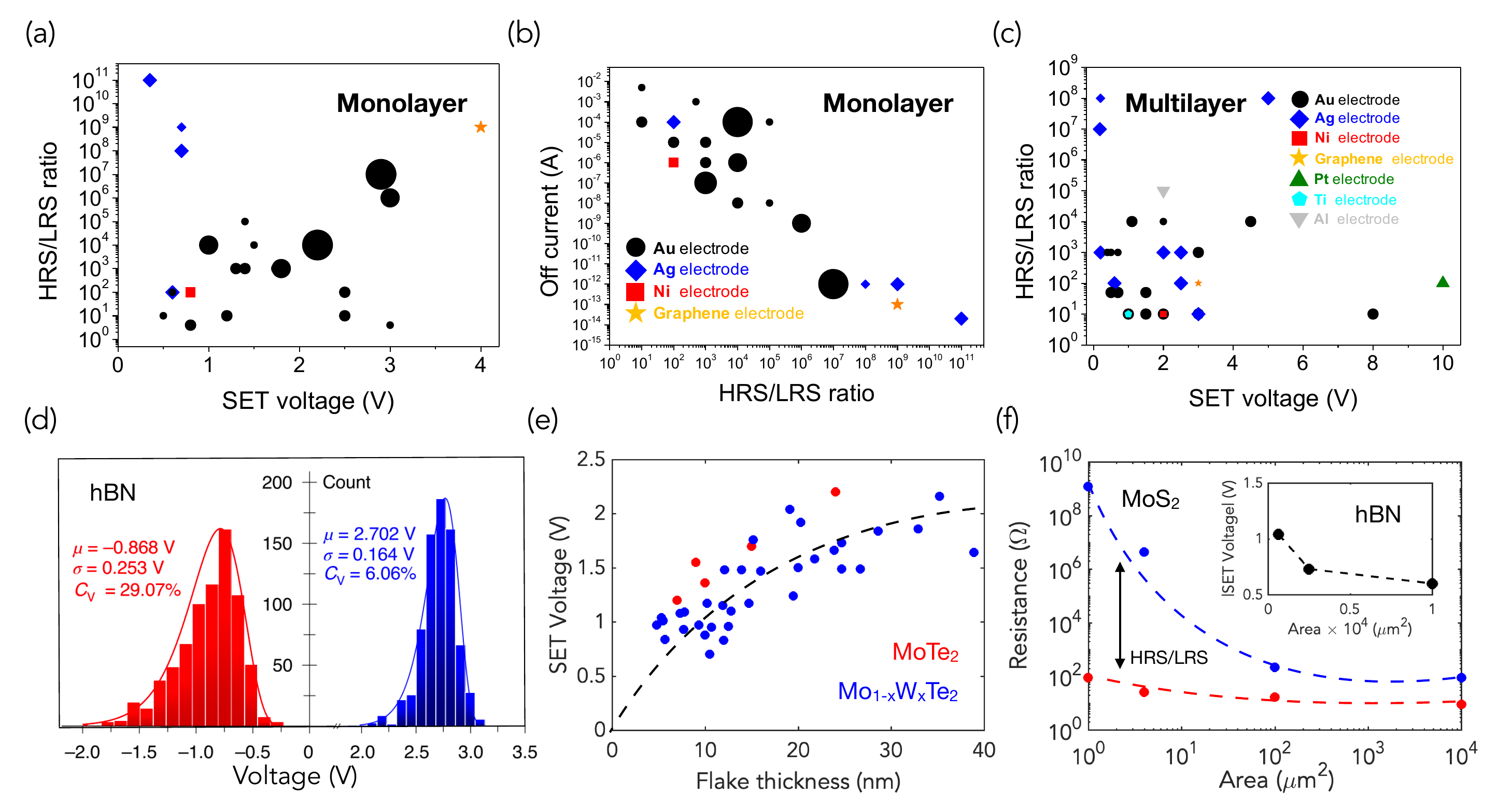}\par
  \caption{\textbf{Device performance and scaling trends for 2DLM RS vertical device stacks made with mono- or multilayer active areas}. (\textbf{a})-(\textbf{c}) show existing device metrics for HRS/LRS ratio, operating voltage (SET Voltage), and HRS current (leakage current) for (\textbf{a})-(\textbf{b}) monolayer and (\textbf{c}) multilayer 2DLM devices. In each case, the OFF-current is read at a voltage of V$_{SET}$/2, and the ON current corresponds to the compliance current used in the measurements. The size of each marker in (\textbf{a-b}) refers to the area of the corresponding device, which ranges from 0.03 $\mu$m$^2$ to 0.20 mm$^2$. The legend (electrode metal) are the same for (\textbf{a})-(\textbf{b}). The data is colored according to the electrode metal. It is tabulated from Refs \cite{Wu2020, Ge2017, Ge2020, Zhao2018-1, Wu2019, Kim2018, Ge2018, Kim2020, Mao2022, Yang2023-2, Yang2024-2} (monolayer) and Refs \cite{He2012, Cheng2015, Das2019, Feng2019, Li2018-8, Kim2018, Qian2016, Zhang2016-5, Zhou2016-7, Han2017, Das2019-2, Rehman2017, Zhang2018, Wang2018, Chen2023-2, Pam2022, Aggarwal2023, Zhuang2023-2, Puglisi2016, Shi2018, Vlkel2023, Lu2021, Li2021-3, Yin2022, Li2021, Hou2023, Zhang_2018, Yang2023} (multilayer). (\textbf{d}) Variation in the RESET and SET Voltages for 48 Au/multilayer hBN/Au device from a single crossbar array, using a compliance current of 1 mA \cite{Chen2020}. (\textbf{e}) Thickness vs. operating voltage for MoTe$_2$ and several Mo$_{1-x}$W$_x$Te devices (x = 0.03 - 0.09), adapted from Ref \cite{Zhang2018}. The black dashed line serves as a guide to the eye. (\textbf{f}) Change in HRS and LRS resistance with device area for Au/monolayer MoS$_2$/Au devices \cite{Ge2017}. The inset shows the effect of area scaling on the SET Voltage for the multilayer hBN stack in Ref \cite{Shi2017}.}
\end{figure*}

\newpage
\newpage

\end{document}